\documentclass{INTERSPEECH2023}

\interspeechcameraready 

\usepackage{multirow}
\usepackage{booktabs}
\usepackage[T1]{fontenc} 
\usepackage[linesnumbered,ruled,vlined]{algorithm2e}
\usepackage[noend]{algpseudocode}
\usepackage{setspace}
\usepackage{xcolor}
\definecolor{mycolor}{RGB}{223,0,124}
\usepackage{hyperref}

\usepackage{stfloats}

\usepackage{caption}
\usepackage{subfigure}

\usepackage{diagbox}

\begin{document}

\title{Robust Training for Speaker Verification against Noisy Labels}

\name{Zhihua Fang$^{1,2}$, Liang He$^{1,2,3,\dagger}$, Hanhan Ma$^{1,2}$, Xiaochen Guo$^{1,2}$, Lin Li$^{4}$
\thanks{
This work was supported by the National Key R\&D Program of China under Grant No. 2022ZD0115801.
Code and data available: \textcolor{mycolor}{\href{https://github.com/PunkMale/OR-Gate}{https://github.com/PunkMale/OR-Gate}}\\
$^\dagger$ Corresponding author.
}}

\address{
  $^1$School of Information Science and Engineering, Xinjiang University, Urumqi 830017, China\\
  $^2$Xinjiang Key Laboratory of Signal Detection and Processing, Urumqi 830017, China\\
  $^3$ Department of Electronic Engineering, Tsinghua University, Beijing 100084, China \\
  $^4$ School of Electronic Science and Engineering, Xiamen University, Xiamen 361005, China
\email{fangzhihua@stu.xju.edu.cn, heliang@mail.tsinghua.edu.cn}
}

\maketitle
 
\begin{abstract}
The deep learning models used for speaker verification rely heavily on large amounts of data and correct labeling.
However, noisy (incorrect) labels often occur, which degrades the performance of the system.
In this paper, we propose a novel two-stage learning method to filter out noisy labels from speaker datasets.
Since a DNN will first fit data with clean labels, we first train the model with all data for several epochs.
Then, based on this model, the model predictions are compared with the labels using our proposed the \emph{OR-Gate} with top-$k$ mechanism to select the data with clean labels and the selected data is used to train the model. This process is iterated until the training is completed.
We have demonstrated the effectiveness of this method in filtering noisy labels through extensive experiments and have achieved excellent performance on the VoxCeleb (1 and 2) with different added noise rates.
\end{abstract}

\section{Introduction}
\label{sec:intro}

In recent years, speaker models based on deep neural networks (e.g, TDNN \cite{tdnn2018icassp}, ResNet \cite{resnet2016CVPR}, and ECAPA-TDNN \cite{ecapa-tdnn2020IS}) have become the mainstream approaches for speaker verification, deriving many variants \cite{pyramid2022icassp,MLP2022icassp,MFA2022IS,dynamicLGL2022IS} with excellent performance.
The success of these models depends on large-scale labelled datasets \cite{voxceleb,cnceleb_2022_SpeechCom}.
Unfortunately, high-quality labelled large-scale data is expensive in practice, and in addition, many large-scale datasets are collected by web crawlers \cite{MS-Celeb-1M}, which inevitably contains noisy labels \cite{inevitably_2019_icassp}.

In the presence of noisy labels in the dataset, the loss function contains wrong labels, which leads to the model to gradient descent towards the wrong direction.
Noisy labels will hinder the model to learn stable speaker features.
Therefore, to learn more accurate speaker features from the data with noisy labels, we must design a robust training strategy for noisy labels.

Researchers have conducted numerous studies on image recognition with noisy labels, and these works can be roughly divided into four categories \cite{survey2022trans}: robust architecture \cite{Yao2019trans,Lee2019ICML}, robust regularization \cite{mixup2018,Wei2021NIPS}, robust loss function \cite{Symmetric2019ICCV,fenglei2020IJCAI}, and sample selection \cite{dividemix2020ICLR,self2020ICLR}.
Learning with noisy labels has been well studied in the field of images, but less in the field of speaker recognition \cite{Li2022,LNCL2021IS,Borgstroem2020PLDA}, so it is necessary to design a speaker recognition method that is robust to noisy labels.

When the training data contains noisy labels, deep neural networks have been observed to fit the training data first with clean labels in the early stages, and then with incorrect labels \cite{early_2022_CVPR,early_2022_NIPS,early_2022_ICML}.
This approach of fitting clean data first and then noisy data is similar to the concept of curriculum learning \cite{Curriculum2009ICML,CL-SR}, which involves gradually fitting training data from easy to difficult in order to improve the performance and generalization ability of neural networks.
Our work has also been inspired by these studies.

In this paper, we propose a novel and effective sample selection method for filtering the noisy labels from speaker datasets.
Our approach is divided into two stages:
Stage I is early learning, we use all the data to train the model for a few epochs, a speaker model with basic discriminatory ability can be obtained. 
Stage II is self-confident learning, which is trained on the model obtained in stage I.
So-called self-confident learning is based on trusting the model and comparing the model predictions with the labels to select clean labels.
Specifically, at each epoch, we compare the sample labels with the predictions of all past epoch models for these samples, and consider the label to be clean (reliable) as long as the match is successful once.
Since this mechanism is similar to the or gate in logic circuits, we name this approach as \emph{OR-Gate}.
All data are divided into clean and noisy data by the \emph{OR-Gate}, and we only use the data with clean labels to update the network parameters, while the data with noisy labels are only forward propagated to get the predictions to be used in the next epoch, and iterate the process until the training is finished.
In summary, the main contributions of this work are as follows:
\begin{itemize}
    \item We propose a two-stage learning method for filtering noisy labels in speaker datasets, and filling the gap of no sample selection method for speaker recognition with noisy labels.
    \item We propose a novel and effective sample selection method called the \emph{OR-Gate}. Since the prediction results of hard samples may not match the labels, we use the top-$k$ mechanism to compare predictions and labels.
    \item We conducted experiments on VoxCeleb 1 and 2 \cite{voxceleb1,voxceleb2} by manually adding various proportions of noisy labels. The experiments demonstrate that our method achieves excellent performance on both the original data and the data with various noise rates. We also evaluate the results of the \emph{OR-Gate} dividing noisy and non-noisy labels which proves the effectiveness of the \emph{OR-Gate} filtering noisy labels.
\end{itemize}

\begin{figure*}[htbp]
\centering
\includegraphics[width=0.8\textwidth,scale=0.4]{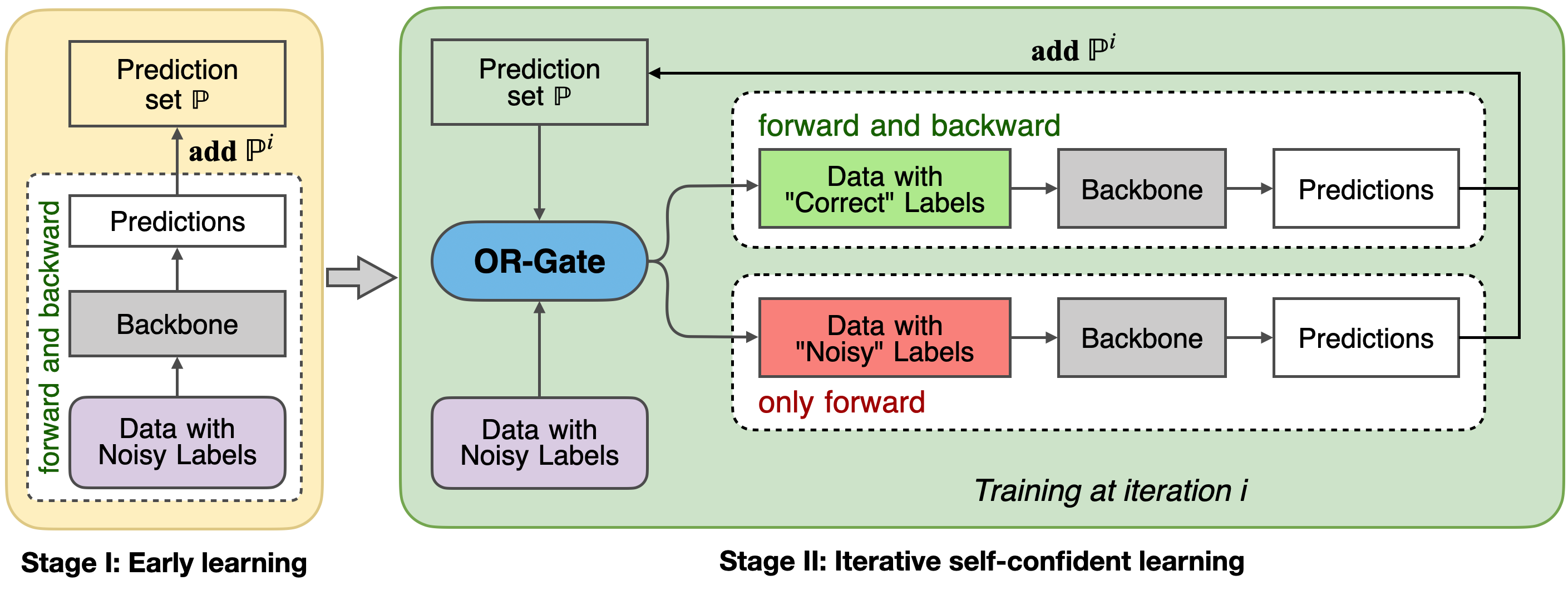}
\caption{
\textbf{The framework of the proposed two-stage learning.}
\textbf{Stage I:} Train the network with all data for a few epochs, and store the model's predictions for each sample at all epochs into the prediction set $\mathbb{P}$.
\textbf{Stage II:} Matching all data with the prediction set $\mathbb{P}$, the data is divided into data with correct labels and data with noisy labels using the OR-Gate (see Figure \ref{fig:orgate} for details). 
Only the data with correct labels are used for training, and for data with noisy labels only the model predictions are output, then the predictions of all samples are added to the prediction set $\mathbb{P}$.
Iterate the above process until the training is completed.
For details, see Algorithm \ref{algo:net}.}
\label{fig:net}
\end{figure*}

\vspace{-0.5em}
\section{Preliminaries}
\label{sec:Pre}

\textbf{Notation.}
For the speaker verification task, let $c$ be the number of speakers and $e$ be a one-hot vector with dimension of $c$.
$D=\left\{\left(x_{i}, y_{i}\right)\right\}_{i=1}^{n}$ denotes the i.i.d. samples and corresponding ground-truth labels, where $n$ is the number of utterances. 
$\widetilde{D}=\left\{\left(x_{i},\widetilde{y}_{i}\right)\right\}_{i=1}^{n}$ is the dataset where the labels are corrupted and the proportion of noisy labels is $\eta$.

\subsection{Problem Setup}
\vspace{-0.3em}
There are two types of noisy labels: symmetric noise and asymmetric noise.
Symmetric noise is generated by randomly flipping the sample labels with equal probability to incorrect labels of other classes, as shown in equation \ref{eq:noisy_label}.
Asymmetric noise is the flipping of sample labels to similar class labels.
The manually added noise labels in the speaker dataset are only applicable to symmetric noise, and the existing speaker recognition with noisy labels only focused on symmetric noise \cite{Li2022,LNCL2021IS}, so this work is based on symmetric noise.
\vspace{-0.2em}
\begin{equation}
    P[\widetilde{y}=e_j \thinspace | \thinspace y=e_i ]=\frac{\eta}{c-1}  \quad \forall \thinspace i \neq j
    \label{eq:noisy_label}
\end{equation}
\vspace{-0.2em}
Let the proportion of noisy labels be $\eta$. The aim is to learn a robust speaker recognition model without knowing the proportion of noisy labels as well as the ground-truth labels.

\subsection{Learning with Noisy Label}
\vspace{-0.3em}
Our work is inspired by SELF \cite{self2020ICLR}, which is a sample selection method of image recognition with noisy labels.
The model attempts to identify correct labels progressively by self-forming ensembles of models and predictions.
It uses a moving-average of ensemble models and predictions to improve filtering decisions, where the ensemble models are actually a vector $\bar{p}$.
The moving-average is updated by the equation:
\begin{equation}
    \bar{P}_k^j=\alpha \bar{P}_k^{j-1}+(1-\alpha) P_k^{j}
\end{equation}
\vspace{-0.2em}
whereby $\bar{P}_t^{j}$ depicts the moving-average of the $t$-th sample at epoch $j$, $\alpha$ is a momentum that represents the confidence in the model prediction as training progresses, $P_t^{j}$ is the model prediction of the $t$-th sample in epoch $j$.
Briefly, $P$ is the combined prediction of the model on the samples. By comparing the sample label $y_i$ with the corresponding $P_i$, we can determine whether the label $y_i$ is clean, and thus filter out data with clean labels for training.

The method is applied to datasets with a small number of labels, while the speaker datasets have a massive number of labels, so it is difficult to correctly classify the ground-truth label samples at the early stage of training, resulting in the inability to filter out a sufficient number of samples for learning, which affects the later training.

\section{The Proposed Approach}
\label{sec:method}

For the problem of noisy labels in the speaker dataset, we propose a two-stage learning approach and maintain a prediction set $\mathbb{P}$ for filtering noisy labels during the whole training process. The details are as follows.

\subsection{Stage I: Early Learning}
Since the number of labels in the speaker dataset is huge, it is difficult for the model to correctly predict the sample labels at the early stage of training, so the early model does not have the ability to filter the samples.
Fortunately, although the neural network will eventually fit all labels, it will first fit the data with clean labels.

Therefore, we first train the model on all data for a few (hyperparameter $w$) epochs to get a model with basic discriminative speaker ability.
This stage is called \emph{Early Learning}, in which the sample predictions for each epoch are stored in the prediction set $\mathbb{P}$.

\subsection{Stage II: Self-Confident Learning}

The model has the basic ability to recognize the speaker after early learning, so the model predictions will be mostly close to the ground-truth labels.
However, since the network parameters fluctuate during the training process, it is difficult to accurately predict the ground-truth labels of the samples in the early phase.
This leads to the inability to select adequate samples for training, which further leads to the inability of the network to learn sufficiently, and this influence will be iterated.

\begin{figure}[htbp] 
	\centering
	\includegraphics[width=0.8\linewidth,scale=1.00]{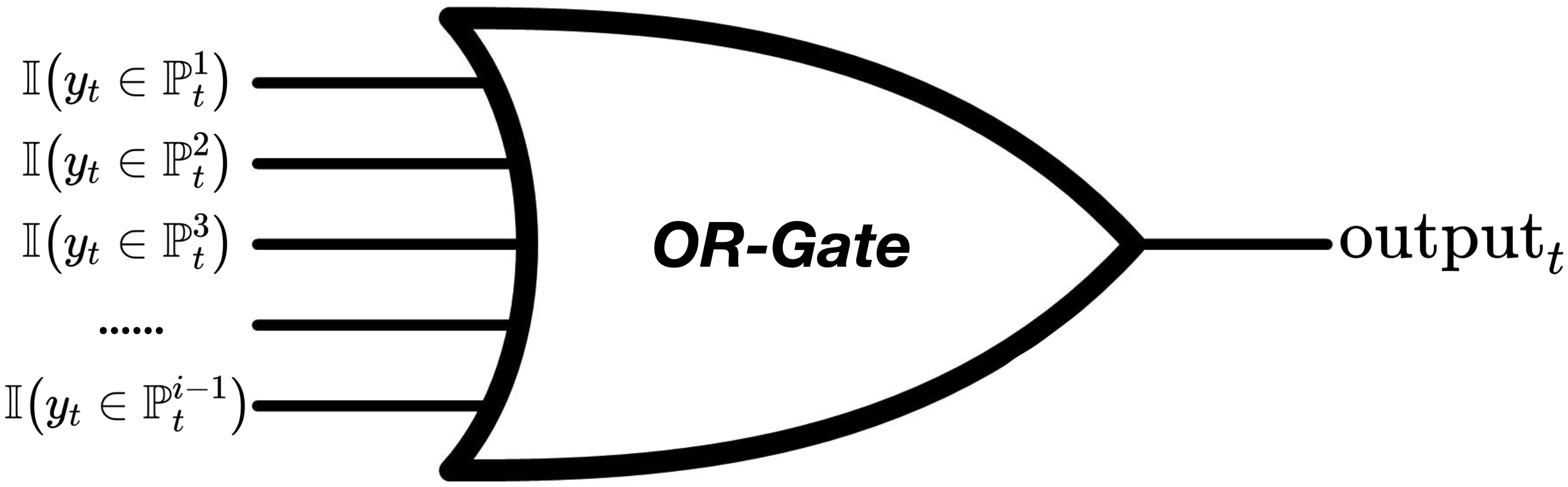}
\caption{
\textbf{The OR-Gate mechanism.}
The picture represents the decision process of the $t$-th sample at the $i$-th epoch.
$\mathbb{I}$ is the indicator function and $\mathbb{P}_t^i$ represents the set of labels for the top-$k$ prediction of the sample $x_t$ at the $i$-th epoch model.
When the $\text{output}_t$ is $1$, the label $y_t$ is correct, otherwise, it is noisy.
}
\label{fig:orgate}
\end{figure}

To solve the above problem, we propose the \emph{OR-Gate}. Similar to the or gate in logic circuits, the sample label is compared with the predictions of the model for this sample in each past epoch, and the label is considered to be clean label as long as there is one successful match. The selected data with clean labels are later used for model training, which is called self-confident learning.

Hard samples often influence the decision boundary in classification tasks, and the model predictions of hard samples generally do not match the ground-truth labels.
To make the \emph{OR-Gate} select more hard samples, we use the top-$k$ mechanism in matching the labels to the predictions, i.e., the matching is considered successful as long as the label $y_i$ exists in the label set $\mathbb{P}_i$ corresponding to the top-$k$ probabilities of the predictions. The details are shown in Figure \ref{fig:orgate}.

\begin{algorithm}[htbp]
    \caption{The proposed two-stage learning algorithm.}
    \label{algo:net}
	\KwIn{training data $\widetilde{D}=\left\{\left(x_{i},\widetilde{y}_{i}\right)\right\}_{i=1}^{n}$, hyperparameters $w$, $k$}
	\KwOut{speaker model}
	Initialize $\mathbb{P}=\emptyset$, $i=0$\\
    \tcp{Stage I: Early Learning}
    \While{$i\text{++} < w$}{
        \For{$(x_t,\widetilde{y}_t)$ \textbf{in} $\widetilde{D}$}{
            $P_t \gets train(\text{Model},x_t,\widetilde{y}_t)$\\
            Construct the label set $\mathbb{P}_t^i$ containing the top-$k$ probabilities of $P_t$\\
            $\mathbb{P}=\mathbb{P}\cup \mathbb{P}_t^i$
        }
    }
	\tcp{Stage II: Self-Confident Learning}
	\While{$i\text{++} < max\_epoch$}{
	    \For{$(x_t,\widetilde{y}_t)$ \textbf{in} $\widetilde{D}$}{
	        \tcp{The OR-Gate mechanism}
		    \eIf{$\sum_{j=1}^{i-1}\mathbb{I}(\widetilde{y}_t \in \mathbb{P}_t^j ) > 0$}{
			    \tcp{Forward and backward}
                $P_t \gets train(\text{Model},x_t,\widetilde{y}_t)$
		    }{
			    \tcp{Forward only}
                $P_t \gets \text{Model}(x)$
		    }
		    Construct the label set $\mathbb{P}_t^i$ containing the top-$k$ probabilities of $P_t$\\
            $\mathbb{P}=\mathbb{P}\cup \mathbb{P}_t^i$
	    }
	}

\end{algorithm}

\section{Experiments}
\label{sec:experiment}

\subsection{Experimental Details}

\textbf{Data.}
We demonstrated the superiority of our proposed method by conducting comprehensive experiments on VoxCeleb1 and 2 with different proportions of noisy labels.
VoxCeleb1 \cite{voxceleb1} contains 1211 speakers and 148,642 utterances, and VoxCeleb2 \cite{voxceleb2} contains 5994 speakers and 109,2009 utterances. Their labels were manually checked and can be considered clean datasets ($\eta=0$).

To verify the noisy label robustness of the method and consider real-world scenarios, we set the noisy label ratio $\eta$ to 0\%, 5\%, 10\%, 20\%, 30\%, and 50\%.
Specifically, for a given noise rate $\eta$, we randomly select the corresponding proportion of samples from VoxCeleb 1 or 2, flip their labels with random equal probability to other error labels, and tag these error samples (these tags are not visible when selecting the samples), and these noise tags can be used in the second stage to evaluate how correctly we partition the data.

\noindent\textbf{Implementation Details.}
For all experiments, we use ResNet34 as the backbone, attentive statistics pooling \cite{asp2018IS} and AM-Softmax \cite{amsoftmax2018SPL} respectively as the pooling layer and loss function.
The input of the network is an 80-dimensional Fbank, the embedding dimension is 512 and the mini-batch size is set to 128.
The network parameters are optimized by the Adam optimizer with an initial learning rate of 0.0002, which is reduced to 40\% every 5 epochs. The backend uses cosine distance to score.

For VoxCeleb 1, we set $w=5$, $k=90$ and trained $80$ epochs.
Since the data volume of VoxCeleb 2 is much larger than that of VoxCeleb 1 and the number of labels is about $5$ times that of VoxCeleb 1, therefore the early learning time of VoxCeleb 2 can be shorter and the value of $k$ can be set larger.
For VoxCeleb 2, we set $w=4$, $k=400$, and trained $60$ epochs.

\noindent\textbf{Performance Metrics.}
We compare \textbf{EER} on test data with clean labels, and since our idea is to filter noisy labels, so we also evaluate the \textbf{precision} (\# \emph{clean labels selected} / \# \emph{of selected labels}) and \textbf{recall} (\# \emph{clean labels selected} / \# \emph{of clean labels in the data}) of identifying noisy labels.

\subsection{Results of comparison experiments}

\textbf{Performance on VoxCeleb 1.}
Table \ref{table:vox1} shows the results of our experiments on VoxCeleb 1.
It can be seen that the performance of the baseline deteriorates rapidly as the noise rate increases, demonstrating that noisy labels have a serious negative impact on speaker verification.
Secondly, when the training data is without noisy labels, our method outperforms the baseline because our method follows the idea of curriculum learning \cite{Curriculum2009ICML,self-paced2010NIPS}.
Since easy samples are easier to predict, they will be selected first by the \emph{OR-Gate}.
This results in the network learning easy samples first and then hard samples, thus outperforming the baseline.
Finally, our method outperforms all other methods when training data with noisy labels.

\begin{table}[!htbp] 
\centering
\caption{EER(\%) comparisons on VoxCeleb1 with different proportions of noisy labels added.}
\resizebox{\linewidth}{!}{
\begin{tabular}{cl|cccccc} 
\toprule 
\multicolumn{2}{c}{Test Data} \vline&
\multicolumn{6}{c}{Vox-O (original test pairs)} \\
\midrule
\multicolumn{2}{c}{$\eta$} \vline & 0\% & 5\% & 10\% & 20\% & 30\% & 50\%  \\
\midrule
\multicolumn{2}{c}{Baseline} \vline & 4.36 & 5.30 & 6.24 & 7.99 & 9.78 & 14.39 \\
\multicolumn{2}{c}{SELF \cite{self2020ICLR} (Re-implemented)} \vline & 5.52 & 5.94 & 5.90 & 6.34 & 7.50 & 11.90 \\
\multirow{3}{*}{LNCL+Sub-AM \cite{LNCL2021IS}}
& + Cosine & 4.64 & 4.83 & 5.11 & 5.45 & 5.56 & 6.32 \\
& + PLDA & 3.92 & 4.45 & 5.02 & 5.47 & 5.96 & 7.29 \\
& + NL-PLDA & \textbf{3.92} & 4.32 & 4.63 & 4.85 & 5.04 & 5.65 \\
\multicolumn{2}{c}{\textbf{OR-Gate}} \vline & 4.08 & \textbf{4.07} & \textbf{4.22} & \textbf{4.28} & \textbf{4.41} & \textbf{5.53} \\
\bottomrule
\label{table:vox1}
\end{tabular}}
\end{table}

\noindent\textbf{Performance on VoxCeleb 2.}
We further conducted experiments on VoxCeleb 2, which contains more speakers and utterances and is relatively difficult to predict, as shown in Table \ref{table:vox2}.
The results show that the proposed method still shows excellent performance for large-scale data containing noisy labels, and has a notable advantage at high noise rates, keeping the EER within 2\% even at a noise rate $ \eta $ of 50\%.

We also observed a special phenomenon when the training data contained few noisy labels, using our method resulted in better performance of the model.
This shows that when the scale of the data is not very different, containing a few noisy labels allows the model to obtain better generalization in early learning and thus to follow the guidelines of curriculum learning (selecting clean samples for training from easy to hard) in the self-confident learning stage.

\begin{table*}[!htbp]
\tabcolsep=0.2cm 
\scriptsize
  \centering
  \caption{EER(\%) comparisons on VoxCeleb2 with different proportions of noisy labels added.}
  \label{table:vox2}
    \resizebox{0.95\textwidth}{!}{
	\begin{tabular}{cl|c c c c c c|c c c c c c}
	\bottomrule
	\multicolumn{2}{c}{Test Data} \vline & \multicolumn{6}{c}{Vox-O (original test pairs)} \vline & \multicolumn{6}{c}{Vox-H (hard test pairs)}\\
	\midrule
	\multicolumn{2}{c}{$\eta$} \vline & 0\% & 5\% & 10\% & 20\% & 30\% & 50\% & 0\% & 5\% & 10\% & 20\% & 30\% & 50\%  \\
	\midrule
	\multicolumn{2}{c}{Baseline} \vline & 1.69 & 1.72 & 1.90 & 2.21 & 2.88 & 4.32 & \textbf{2.89} & 3.11 & 3.25 & 3.64 & 4.42 & 6.74  \\
    \multirow{3}{*}{LNCL+Sub-AM \cite{LNCL2021IS}} & + Cosine & \textbf{1.63} & \textbf{1.63} & 1.64 & 1.69 & 1.80 & 2.25 & 3.02 & 3.05 & 3.06 & 3.26 & 3.26 & 3.77  \\
    & + PLDA & 1.71 & 1.78 & 1.81 & 2.05 & 2.15 & 2.86 & 3.03 & 3.10 & 3.15 & 3.57 & 3.76 & 4.76  \\ 
    & + NL-PLDA & 1.70 & 1.72 & 1.73 & 1.75 & 1.77 & 2.19 & 3.02 & 3.07 & 3.10 & 3.20 & 3.33 & 3.58  \\
    \multicolumn{2}{c}{\textbf{OR-Gate}} \vline & 1.64 & 1.65 & \textbf{1.62} & \textbf{1.67} & \textbf{1.72} & \textbf{1.97} & 2.93 & \textbf{2.71} & \textbf{2.77} & \textbf{2.96} & \textbf{2.87} & \textbf{3.11}  \\
    \bottomrule
	\end{tabular}}
\end{table*}

\subsection{Efficacy of detecting clean samples.}

Although the EER results on noise-added VoxCeleb 1 and 2 demonstrate the validity of the \emph{OR-Gate}, it is also meaningful to evaluate the precision and recall of the selected samples because of the sample selection method we used.

\noindent\textbf{Precision analysis.}
Precision can reflect the ability of our method to filter noisy labels.
As can be seen from Table \ref{table:recall}, the highest precision (almost $1$) was achieved on both datasets after early learning to divide the data, demonstrating that early learning to first fit the data with clean labels is what makes the data filtered after early learning almost free of noisy labels.
Although the precision decreases during the self-confident learning period, the precision remains above 98\% throughout the training period until the end, and the tiny percentage of noisy labels has minimal impact on the model.

\noindent\textbf{Recall analysis.}
Recall reflects the ability of our method to select clean samples.
As can be seen from Table \ref{table:recall}, the \emph{OR-Gate} can select a large portion of clean data after early learning, except for the case of the small dataset (VoxCeleb 1) with high noise rate (50\%).
This proves that a model with basic speaker recognition capability can be obtained after early learning, which proves that our idea is right.
When the noise rate is below 50\%, the \emph{OR-Gate} selects more than 98\% of clean samples by self-confident learning, and when the noise rate is as high as 50\%, the \emph{OR-Gate} also selects the majority of clean samples.
The above data demonstrate the \emph{OR-Gate} can select as many clean samples as possible while filtering noisy labels.

\begin{table*}[!htbp]
\scriptsize
  \centering
  \caption{Precision and Recall of selecting clean labels at different periods of training.}
  \label{table:recall}
    \resizebox{0.95\textwidth}{!}{
	\begin{tabular}{c|c c c c c c|c c c c c c}
	\bottomrule
	Metric & \multicolumn{6}{c}{Precision} \vline & \multicolumn{6}{c}{Recall}\\
	\midrule
	$\eta$ & 0\% & 5\% & 10\% & 20\% & 30\% & 50\% & 0\% & 5\% & 10\% & 20\% & 30\% & 50\%  \\
	\midrule
	\midrule
	\multicolumn{13}{c}{VoxCeleb 1} \\
	\midrule
	$\text{epoch}_{6}$ & \textbf{1.0000} & \textbf{0.9999} & \textbf{0.9999} & \textbf{0.9999} & \textbf{0.9999} & \textbf{1.0000} & 0.9523 & 0.9373 & 0.9161 & 0.8298 & 0.7018 & 0.3054  \\
    $\text{epoch}_{10}$ & 1.0000 & 0.9998 & 0.9997 & 0.9996 & 0.9992 & 0.9974 & 0.9857 & 0.9841 & 0.9800 & 0.9603 & 0.9250 & 0.6810 \\
    $\text{epoch}_{20}$ & 1.0000 & 0.9997 & 0.9995 & 0.9989 & 0.9980 & 0.9928 & 0.9942 & 0.9940 & 0.9933 & 0.9897 & 0.9780 & 0.8123 \\
    $\text{epoch}_{50}$ & 1.0000 & 0.9995 & 0.9990 & 0.9981 & 0.9963 & 0.9876 & 0.9975 & 0.9975 & 0.9971 & 0.9961 & 0.9916 & 0.8586 \\
    $\text{epoch}_{80}$ & 1.0000 & 0.9995 & 0.9988 & 0.9976 & 0.9953 & 0.9847 & \textbf{0.9980} & \textbf{0.9982} & \textbf{0.9978} & \textbf{0.9969} & \textbf{0.9929} & \textbf{0.8598} \\
    \midrule
    \multicolumn{13}{c}{VoxCeleb 2} \\
    \midrule
    $\text{epoch}_{5}$ & \textbf{1.0000} & \textbf{0.9999} & \textbf{0.9999} & \textbf{0.9998} & \textbf{0.9999} & \textbf{0.9999} & 0.9774 & 0.9743 & 0.9706 & 0.9545 & 0.9202 & 0.7093 \\
    $\text{epoch}_{10}$ & 1.0000 & 0.9999 & 0.9997 & 0.9994 & 0.9992 & 0.9979 & 0.9842 & 0.9837 & 0.9829 & 0.9806 & 0.9762 & 0.9369 \\
    $\text{epoch}_{20}$ & 1.0000 & 0.9998 & 0.9996 & 0.9991 & 0.9986 & 0.9961 & 0.9865 & 0.9862 & 0.9858 & 0.9845 & 0.9818 & 0.9499 \\
    $\text{epoch}_{40}$ & 1.0000 & 0.9997 & 0.9995 & 0.9988 & 0.9980 & 0.9945 & 0.9880 & 0.9878 & 0.9876 & 0.9866 & 0.9841 & 0.9552 \\
    $\text{epoch}_{60}$ & 1.0000 & 0.9997 & 0.9994 & 0.9986 & 0.9977 & 0.9937 & \textbf{0.9888} & \textbf{0.9886} & \textbf{0.9884} & \textbf{0.9873} & \textbf{0.9850} & \textbf{0.9569} \\
    \bottomrule
	\end{tabular}}
\end{table*}

\subsection{Ablation experiments}
We also conducted ablation experiments.
Table \ref{table:ablation} shows that the performance of the model without early learning is much worse than the baseline, which indicates that early learning is essential in our approach. It is only through early learning that a speaker model with basic recognition ability can be obtained for "self-confident" learning of Stage II.
In addition, we can see that adding the top-$k$ mechanism to the \emph{OR-Gate} can improve the performance more than not adding it, which indicates that the top-$k$ mechanism can help the model select more hard samples to improve the classification ability of the model and get higher quality speaker embedding.

\begin{table}[!htbp] 
\centering
\caption{EER (\%) comparisons of the ablation experiments on VoxCeleb 1.}
\label{table:ablation}
\resizebox{\linewidth}{!}{
\begin{tabular}{l|cccccc} 
\toprule 
\multicolumn{1}{c}{Test Data} \vline & \multicolumn{6}{c}{Vox-O (original test pairs)} \\
\midrule
\multicolumn{1}{c}{$\eta$} \vline & 0\% & 5\% & 10\% & 20\% & 30\% & 50\%  \\
\midrule
Baseline & 4.36 & 5.30 & 6.24 & 7.99 & 9.78 & 14.39 \\
OR-Gate w/o early learning & 5.99 & 6.27 & 6.87 & 8.37 & 13.05 & 25.83 \\
OR-Gate w/o top-$k$ mechanism & 5.21  & 4.81  & 5.15  & 5.58  & 6.17  & 8.55 \\
OR-Gate & \textbf{4.08} & \textbf{4.07} & \textbf{4.22} & \textbf{4.28} & \textbf{4.41} & \textbf{5.53}\\
\bottomrule
\end{tabular}}
\end{table}

\vspace{-0.5em}
\section{Conclusions}
\label{sec:conclusion}

In this paper, we propose a novel and easy-to-implement framework for filtering noisy labels in speaker datasets.
Specifically, a model with basic speaker discrimination ability is first obtained by early learning, and then self-confident learning is conducted based on this model, where the network is trained using our proposed the \emph{OR-Gate} with the top-$k$ mechanism to select clean data.
Excellent performance was achieved on both VoxCeleb 1 and 2 with various noise ratios added.

\bibliographystyle{IEEEtran}
\bibliography{Interspeech}

\end{document}